\documentclass[prd,twocolumn,showpacs,amsmath,amssymb,floatfix,nofootinbib]{revtex4}
\usepackage[colorlinks=true, citecolor=blue, linkcolor=blue, urlcolor=blue]{hyperref}
\usepackage{amsfonts,bm,color,dcolumn,xcolor,graphicx,amsmath,multirow}
\usepackage{amssymb,latexsym,subfigure,threeparttable,multirow,makecell,tabularx,setspace,textcomp}
\usepackage[mathscr,scaled=1.15]{urwchancal}
\DeclareFontFamily{OT1}{pzc}{}
\DeclareFontShape{OT1}{pzc}{m}{it}{<->s*[1.15]pzcmi7t}{}
\DeclareMathAlphabet{\mathpzc}{OT1}{pzc}{m}{it}

\allowdisplaybreaks

\newcommand{\DS}[1]{/\!\!\!#1}

\newcommand{\q}{\mathpzc q^{a_0}}

\begin{document}
\title{Searching for $a_0(980)$-meson parton distribution function}
\author{Zai-Hui Wu}
\author{Hai-Bing Fu}
\email{fuhb@cqu.edu.cn (Corresponding author)}
\author{Tao Zhong}
\email{zhongtao1219@sina.com}
\author{Yu Chen}
\author{Ya-Hong Dai}
\address{Department of Physics, Guizhou Minzu University, Guiyang 550025, P.R. China}

\date{\today}
\begin{abstract}
In this paper, we calculate the scalar $a_0(980)$-meson leading-twist wave function by using light-cone harmonic oscillator model (LCHO). In which the model parameters are determined by fitting the $\xi$-moments $\langle\xi_{a_0}^n\rangle_\zeta$ of its light-cone distribution amplitudes. Then, the $a_0(980)$-meson leading-twist light-cone distribution amplitudes with three different scales $\zeta= (1.0, 2.0, 5.2)~{\rm GeV}$ are given. After constructing the relationship between $a_0(980)$-meson leading-twist parton distribution functions/valence quark distribution function and its LCHO wave function, we exhibit the $\mathpzc q^{a_0}(x,\zeta)$ and $x\mathpzc q^{a_0}(x,\zeta)$ with different scales. Furthermore, we also calculate the Mellin moments of the $a_0(980)$-meson's valence quark distribution function  $\langle x^n \mathpzc q^{a_0}\rangle_\zeta$ with $n = (1,2,3)$, i.e. $\langle x \mathpzc q^{a_0}\rangle_{\zeta_5} = 0.026$, $\langle x^2 \mathpzc q^{a_0}\rangle_{\zeta_5} = 0.017$ and $\langle x^3 \mathpzc q^{a_0}\rangle_{\zeta_5} = 0.012$. Finally, the scale evolution for the ratio of the Mellin moments $\mathpzc x^n_{\,a_0}(\zeta,\zeta_k)$ are presented.
\end{abstract}

\maketitle

\textit{1. Introduction} --- The exploration of quark-gluon structure in hadrons is a cross-cutting edge issue between particle physics and medium-high energy nuclear physics in recent years. In which quarks and gluons, called partons, are the fundamental degrees of freedom of quantum chromodynamics (QCD). Although the parton can not be directly observed, the QCD factorization theorem allows one to express the information of the parton inside the nucleon in terms of nonperturbative functions~\cite{Collins:1989gx}. At the same time, the parton distribution function (PDF) is considered to be the most important nonperturbative function, which plays an important role in describing the nonperturbative QCD for the internal structure of hadronic bound states~\cite{Berger:1979du}. In addition, it also gives the probability of finding quarks and gluons inside a hadron. In the infinite momentum coordinate system~\cite{Gribov:1973jg, Bjorken:1969ja, Feynman:1969ej, Bjorken:1968dy}, the PDFs are used to describe the one-dimensional momentum distributions of quarks and gluons. Therefore, the internal structure of hadrons can be studied by calculating the meson's PDF.

The PDF constitute the basic limit of Higgs boson representation in terms of coupling, which is the main system for standard model (SM) measurement such as $W$-boson mass. And it is also still the largest uncertainty outside the production of SM heavy particles, so it has important phenomenological value. The MMHT\cite{Harland-Lang:2014zoa}, CT~\cite{Dulat:2015mca}, NNPDF~\cite{NNPDF:2017mvq}, HERAPDF~\cite{Alekhin:2017kpj}, and JAM~\cite{Ethier:2017zbq} have made substantial efforts to determine PDFs and their uncertainties. The pion deemed to the lightest bound state of QCD and kaon have been predicted by many theoretical calculations, chiral-quark model~\cite{Nam:2012vm,Watanabe:2016lto, Watanabe:2017pvl}, Nambu-Jona-Lasinio model~\cite{Hutauruk:2016sug}, light-front holographic QCD (LHFQCD)~\cite{deTeramond:2018ecg,Chang:2020kjj,Watanabe:2019zny,Lan:2020fno}, light front quantization ~\cite{Lan:2019vui,Lan:2019rba,Lan:2020hyb}, maximum entropy method~\cite{Han:2018wsw, Han:2020vjp}, Dyson-Schwinger equations (DSEs)~\cite{Chang:2014lva, Chang:2014gga, Chang:2021utv, Chen:2016sno,Shi:2018mcb, Bednar:2018mtf, Ding:2019lwe, Freese:2021zne, Cui:2021mom, Cui:2020tdf, Cui:2022bxn} and lattice QCD~\cite{Gao:2022iex, Gao:2020ito, Gao:2021dbh, Sufian:2020vzb, Lin:2020ssv, Joo:2019bzr, Shugert:2020tgq, Izubuchi:2019lyk, Sufian:2019bol, Zhang:2018nsy} for the valence quark PDF. In most cases, the light front wave function (LFWF) is adopted to predict the PDFs. The LFWF is obtained by calculating the Bethe-Salpeter wave functions by the covariant DSEs and projecting the quantity in the Bethe-Salpeter wave functions to the light front in Ref.~\cite{Cui:2022bxn}. At the same time, other models have also analyzed LFWF, such as LHFQCD~\cite{deTeramond:2018ecg, Chang:2020kjj, Watanabe:2019zny,Lan:2020fno} and BLFQ~\cite{Lan:2019vui, Lan:2019rba, Lan:2020hyb}. Besides, there are few papers on the PDFs of the scalar meson $a_0(980)$ with the same quark $q\bar{q}$ state, which motivates us a research motivation.

One of the earliest predictions for $J = 0$ meson valence-quark distribution function for large-$x$ behaviour within the QCD improved parton model~\cite{Ezawa:1974wm, Farrar:1975yb, Holt:2010vj}, has the following expression:
\begin{eqnarray}
\mathpzc q^M(x;\zeta) \overset{x \simeq 1}{=} \mathpzc c(\zeta)(1-x)^{\beta_\zeta},~{\beta_\zeta} =2,
\end{eqnarray}
where $\mathpzc c(\zeta)$ is independent of $x$ and the $\zeta$ stands for the resolving scale. Since the meson's PDF can be obtained directly from its wave function (WF), a more accurate $a_0(980)$-meson WF is crucial for us to determine its PDF. The light-cone harmonic oscillator model (LCHO) for the light or heavy meson WF is mainly based on the Brodsky-Huang-Lepage (BHL) prescription~\cite{BHL}, which have been used in many cases~\cite{Zhong:2014jla, Zhong:2021epq, Wu:2012kw, Huang:2013gra, Huang:2013yya, Zhong:2015nxa, Huang:1994dy, Huang:2004fn, Wu:2011gf}. For this model, the total WF can be separated into spin-space WF $\chi_{M}^{\lambda_1\lambda_2}(x,{\bf k}_\bot)$ and spatial wave function $\psi^R_{M}(x,{\bf k}_\bot)$. The spatial WF is divided into $x$-dependence part and ${\bf k}_\bot$-dependence part for calculation. The ${\bf k}_\bot$-dependence part derives from the approximate bound-state solution and the $x$-dependence part $\varphi_{M}(x)$ can be expanded in Gegenbauer polynomials.

Furthermore, the meson light-cone distribution amplitudes (LCDAs) can also been related to its WF, which leads to the indirect relationship between meson LCDAs and PDF. In many applications of LCDAs, one usually takes a truncated form to determine DAs, involving only the first few terms of the Gegenbauer expansion series. With the increase of $n$, there will exist dimensional anomalies which lead to spurious oscillations. In addition, one of the most important factors is the unreliability of the higher-order Gegenbauer moments. In order to improve this phenomenon, one can adopt LCHO model to deal with the meson LCDAs. In our previous works~\cite{Zhong:2021epq, Hu:2021lkl}, meson's the leading-twist LCDAs are studied by using the LCHO model, and then the model parameters are determined by fitting moments with the least squares method. Therefore, the $a_0(980)$-meson PDFs will be studied based on BHL prescription in this paper.\\

\textit{2. Theoretical Framework} --- If one wants to use the typical probability expression of quantum mechanics to describe the measurable properties of a given hadron, the first thing one needs to do is finding the WF. Each element of the WF fock-space decomposition represents the probability amplitude of finding $n$ components in the hadron. However, the PDFs describe the longitudinal momentum distribution parton of hadron. To derive the $a_0(980)$-meson leading-twist PDF, the following expression can be used~\cite{Cui:2020tdf}
\begin{eqnarray}
\mathpzc q^{a_0}(x,\zeta) = \int_{|{\bf k}_\bot|^2 \leq \zeta^2} d^2{\bf k}_\bot |\psi_{a_0}(x,{\bf k}_\bot)|^2.
\label{PDF_WF}
\end{eqnarray}
where $\psi_{a_0}(x,\textbf{k}_\bot)$ is WF and $\zeta$ stands for the scale. In order to calculate the PDF, the connection between distribution amplitudes and distribution function can be established to achieve the purpose. Exploiting this relationship, one can predict $a_0(980)$-meson leading-twist PDFs with the LCHO model based on BHL description \cite{BHL}. The LCHO model of the $a_0(980)$-meson leading-twist WF is denoted by
\begin{align}
&\psi_{a_0}(x,{\bf k}_\bot)= \sum_{\lambda_1\lambda_2} \chi_{a_0}^{\lambda_1\lambda_2}(x,{\bf k}_\bot) \psi^R_{a_0}(x,{\bf k}_\bot),
\label{WF_full}
\end{align}
with ${\bf k}_\bot$ is transverse momentum. What's more, $\lambda_1$ and $\lambda_2$ are the helicities of the two constituent quark. The spin-space WF $\chi_{a_0}^{\lambda_1 \lambda_2} (x, {\bf k}_\bot)$ comes from the Wigner-Melosh rotation. The different forms for $\lambda_1\lambda_2$ can also be found in Ref.~\cite{Huang:1994dy}. Thus the sum of the spin-space WF have the following form
\begin{align}
&\sum_{\lambda_1\lambda_2}\chi_{a_0}^{\lambda_1\lambda_2}(x,{\bf k}_\bot) =\frac{m_q^2}{\sqrt{{\bf k}^2_\bot + m_q^2}},
\end{align}
with $m_q=m_u=m_d$. On the other hand, the BHL description~\cite{Guo:1991eb, Huang:1994dy} proposed the assumption that the valence Fock WF depends only on the energy variable $\epsilon$ outside the shell. At the same time, the connection between the equal-time WF in the rest frame and the light-cone wave function in the infinite frame by equating the energy propagator $\epsilon =M^2-(\sum_{i=1}^{n} k_i)^2$ is proposed. The propagators in different frames are as follws
\begin{align}
&\epsilon_1 =M^2-\left(\sum\limits _{i=1}^{n} q^0_i\right)^2, &&\sum\limits _{i=1}^n {\bf q}^i=0
 \nonumber\\
&\epsilon_2 =M^2-\bigg(\sum\limits _{i=1}^{n}\frac{{\bf k}^2_{\bot i}+m^2_i}{x_i}\bigg),&&\sum\limits _{i=1}^{n} {\bf k}^2_{\bot i}=0,
 \nonumber\\
& && \sum\limits_{i=1}^{n} x^i=1.
\end{align}
For the two-particle system, one can get
\begin{align}
{\bf q} \leftrightarrow \frac{{\bf k}^2_{\bot i}+m^2_q}{4x(1-x)}-m^2_q
\end{align}
where $q^0_1= q^0_2$. Besides, there is approximate connection between $\psi_{\rm CM}({\bf q})$ and $\psi_{\rm LC}(x,{\bf k})$,
\begin{align}
\psi_{\rm CM}({\bf q}^2) \leftrightarrow \psi_{\rm LC}\bigg(\frac{{\bf k}^2_{\bot} + m^2_q}{4x(1-x)}-m^2_q\bigg)
\label{CM}
\end{align}
Based on the approximation for bound state solution of the meson quark model, the WF of the harmonic oscillator model in the rest frame is expressed as
\begin{align}
\psi_{\rm CM}({\bf q}^2)=A  \exp\bigg(-\frac{{\bf q}^2}{2\beta^2}\bigg)
\label{CM1}
\end{align}
Combining Eqs.~\eqref{CM} and ~\eqref{CM1}, the spatial WF is
\begin{align}
&\psi^R_{a_0}(x,{\bf k}_\bot)= A_{a_0} \varphi_{a_0}(x) \exp \left[ -\frac{{\bf k}^2_\bot + m_q^2}{8\beta_{a_0}^2 x\bar x} \right].
\label{WF_spatial}
\end{align}
where $A_{a_0}$ is the normalization constant and $\varphi_{a_0}(x) = (x\bar x)^{\alpha_{a_0}} C_1^{3/2}(2x-1)$. Thus, one can get the $a_0(980)$-meson leading-twist WF
\begin{align}
\psi_{a_0}(x,{\bf k}_\bot)=\frac{m_q^2 A_{a_0} \varphi_{a_0}(x)}{\sqrt{{\bf k}^2_\bot + m_q^2}}\exp \left[ -\frac{{\bf k}^2_\bot + m_q^2}{8\beta_{a_0}^2 x\bar x} \right].
\end{align}
Furthermore, the $a_0(980)$-meson leading-twist valence-quark distribution function can be obtained by integrating over the squared transverse momentum, i.e. Eq.~\eqref{PDF_WF}, which leads to the following formula
\begin{align}
&\mathpzc q^{a_0}(x,\zeta)=\frac{ A^2_{a_0} m_q^2 (1-2x)^2(x\bar{x})^{2\alpha_{a_0}}}{512\pi^5}
\nonumber\\
&\quad\times \left\{ \textrm{Ei}\left[ \sqrt{\frac{m_q^2+\zeta^2 }{4\beta_{a_0}^2 x\bar x}} \right] - \textrm{Ei}\left[ \sqrt{\frac{m_q^2}{4\beta_{a_0}^2 x\bar x}} \right] \right\},
\label{PDF_model}
\end{align}
with $\textrm{Ei}(x)=-\int^\infty _{-z} \frac{e^{-t}}{t}dt $. It guarantees the relationship
\begin{align}
\int_0^1 dx \mathpzc q^{a_0}(x,\zeta) = 1
\end{align}
which is in agreement with the cases of pion and kaon. For further research, the process-independent effective charge is used to redesign the process-dependent-charge alternative and implement evolution to integrate the one-loop Dokshitzer-Gribov-Lipatov-Altarelli-Parisi (DGLAP) equations with describing the evolution of quarks or gluons fragmenting into hadrons and identifying only a hadron at a time. The relevant explanation can be found in Ref.~\cite{Cui:2020tdf}. Using this process, one can get the Mellin moments of $a_0(980)$-meson's valence-quark distribution function:
\begin{align}
\langle x^n \mathpzc q^{a_0}\rangle_{\zeta}=\int^1_0 x^n \mathpzc q^{a_0}(x,\zeta)dx
\end{align}
Then, the ratio of Mellin moments is also a point of interest, and its expression is as follows
\begin{align}
\mathpzc x^n_{~a_0}(\zeta,\zeta_k)=\frac{\langle x^n \mathpzc q^{a_0}\rangle_\zeta}{\langle x^n \mathpzc q^{a_0} \rangle_{\zeta_k}}
\end{align}

Another significant physical quantity associated with $a_0(980)$-meson PDF is its LCDA. The relationship between the $a_0(980)$-meson leading-twist LCDA and the WF is
\begin{eqnarray}
\phi_{a_0}(x,\zeta) = \int_{|{\bf k}_\bot|^2 \leq \zeta^2} \frac{d^2{\bf k}_\bot}{16\pi^3} \psi_{a_0}(x,{\bf k}_\bot).
\label{DA_WF}
\end{eqnarray}
After integrating over the squared transverse momentum, one can get the LCDA formula
\begin{align}
&\phi_{a_0}(x,\zeta)=\frac{ A_{a_0} m_q \beta_{a_0}}{4\sqrt{2}\pi^{3/2}} \sqrt{x\bar x} \varphi_{a_0}(x)
\nonumber\\
&\quad\times \left\{ \textrm{Erf}\left[ \sqrt{\frac{m_q^2+ \zeta^2 }{8\beta_{a_0}^2 x\bar x}} \right] - \textrm{Erf}\left[ \sqrt{\frac{m_q^2}{8\beta_{a_0}^2 x\bar x}} \right] \right\},
\label{DA_model}
\end{align}
where ${\rm Erf}(x) = 2\int^x_0 e^{-t^2} dx/\sqrt\pi$ is the error function. In order to determine the free model parameters $A_{a_0}$, $\beta_{a_0}$ and $\alpha_{a_0}$, we should use the $\xi$-moments of the $a_0(980)$-meson leading-twist LCDA, which has the following definition
\begin{align}
\langle\xi^n_{a_0}\rangle_\zeta=\int^1_0 \xi^n \phi_{a_0}(x,\zeta)dx
\label{xin}
\end{align}
On the other hand, the $\xi$-moments can be calculated by QCD sum rule approach within background field theory (BFTSR). The two-point correlation function is taken as
\begin{eqnarray}
\Pi_{a_0}^{(n,0)} = i\int d^4x e^{iq\cdot x}\langle0|T\{J_n^V(x), J_0^{S,\dag}(0)\}|0\rangle,
\end{eqnarray}
with $n$ will take the odd numbers, while the even order will vanish due to the $G$-parity. The currents are $J^V_n(x) = \bar q_1(x) \DS z (i z\cdot \tensor{D})^n q_2(x)$ and $J^S_0(0) = \bar q_1(0)q_2(0)$. The detail calculation process for the $\xi$-moments are given in our recent paper~\cite{Wu:2022qqx}.

Then, one can adopt the least squares method to fit $\xi$-moments $\langle \xi^n_{a_0} \rangle_\zeta$ in determining the free model parameters. The purpose of the least squares method is to obtain the optimal value of the fitting parameter $\theta$ through minimizing the likelihood function
\begin{align}
\chi^2(\theta)=\sum\limits_{i=1}^{N}\frac{y_i-\mu(x_i,\theta)}{\sigma^2_i}
\end{align}
$\mu(x_i,\theta)$ is the $a_0(980)$-meson $\xi$-moments $\langle\xi^n_{a_0}\rangle_\zeta$ of combining Eqs.~\eqref{DA_model} and \eqref{xin}. The value of $y_i$ and its variance $\sigma_i$ are defined the value of $\xi$-moments calculated by QCD sum rule. And beyond that, making use of the probability density function $f(y,n_d)=(1/\Gamma(\frac{n_d}{2})2^{n_d/2})y^{n_d/2-1}e^{-\frac{y}{2}}$ of $\chi^2$, one can get the goodness of fit with the following probability $P_{\chi^2}$
\begin{align}
P_{\chi^2}=\int^1_0(f_y;n_d)dy
\end{align}
with $P_{\chi^2}\in(0,1)$. The closer the goodness of fit is to $1$, the better the parameters are obtained. Incorporating the effect of scale $\zeta$, according to the renormalization group equations (RGE) of Gegenbauer moments of the $a_0(980)$-meson leading-twist LCDA,
\begin{eqnarray}
a_n^{a_0}(\zeta) &=& a_n^{a_0}(\zeta_0) E_n(\zeta, \zeta_0),
\label{Eq:anRGE}
\end{eqnarray}
where $E_n(\zeta, \zeta_0) = [\alpha_s(\zeta)/\alpha_s(\zeta_0)]^{-(\gamma_n^{(0)}+4)/b}$ and the coefficient $b = (33-2n_f)/3$~\cite{Cheng:2005nb}. $n_f$ is the number of active quark flavors. The $\zeta_0$ and $\zeta$ are considered as the initial scale and the running scale. Here we make a notation that the $\xi$-moments can translate into $a_n$-moment directly. The one-loop anomalous dimensions is
\begin{eqnarray}
\gamma_n^{(0)} = C_F\bigg[1-\frac{2}{(n+1)(n+2)}+4\sum^{n+1}_{j=2}\frac1j\bigg]
\end{eqnarray}
with $C_F = 4/3$. Then, one can gain the $\xi$-moments at the arbitrary scales $\zeta$. Finally, after fitting the moments $\langle\xi^n_{a_0}\rangle_\zeta$ with the least squares method, we can predict the $a_0(980)$-meson valence quark distribution function, Mellin moments $\langle x^n \mathpzc q^{a_0}\rangle_{\zeta}$ and ratio $\mathpzc x^n_{\,a_0}(\zeta,\zeta_k)$.\\

\begin{table}[b]
\centering
\renewcommand\arraystretch{1.3}
\small
\caption{The $a_0(980)$-meson leading-twist LCDA moments $\langle\xi^n_{a_0}\rangle_\zeta$ at scale $\zeta=(1.0,2.0,5.2)~{\rm GeV}$.}
\begin{tabular}{c  c  c  c  c  c}
\hline
                                         & $\zeta_0$            & $\zeta_2$            & $\zeta_5$ \\ \hline
$\langle\xi^1_{a_0}\rangle_\zeta$ & $-0.307\pm0.043$ & $-0.212\pm0.030$ & $-0.158\pm0.022$ \\
$\langle\xi^3_{a_0}\rangle_\zeta$ & $-0.181\pm0.034$ & $-0.091\pm0.017$ & $-0.048\pm0.009$ \\
$\langle\xi^5_{a_0}\rangle_\zeta$ & $-0.078\pm0.028$ & $-0.055\pm0.017$ & $-0.039\pm0.011$ \\
$\langle\xi^7_{a_0}\rangle_\zeta$ & $-0.049\pm0.026$ & $-0.035\pm0.015$ & $-0.025\pm0.010$ \\
$\langle\xi^9_{a_0}\rangle_\zeta$ & $-0.036\pm0.024$ & $-0.011\pm0.012$ & $~0.000\pm0.006$ \\
\hline
\end{tabular}
\label{Tab:moments}
\end{table}

\textit{3. Numerical analysis} --- To do the numerical analysis, the following input parameters are used. The mass of $a_0(980)$-meson is taken as $m_{a_0} = 0.980\pm0.020~{\rm GeV}$. The current light quark-mass, charm quark mass, the values of the non-perturbative vacuum condensates, the continuum threshold $s_0$ and the corresponding Borel windows used in $\xi$-moments BFTSR are consistent with our previous work~\cite{Wu:2022qqx}. The light quark constitute mass is taken as $m_q = 250 ~{\rm MeV}$. In this paper, we take three typical scales, the initial scales $\zeta_0 = 1.0~{\rm GeV}$, the processes scale $\zeta_2 = 2.0~{\rm GeV}$ and the scale for $\pi$-nucleon Drell-Yan experiment~\cite{Conway:1989fs} or the E615 experiment $\zeta_5 = 5.2 ~{\rm GeV}$, which agree with the pion cases~\cite{Cui:2020tdf}.
\begin{figure}[t]
\centering
\includegraphics[width=0.45\textwidth]{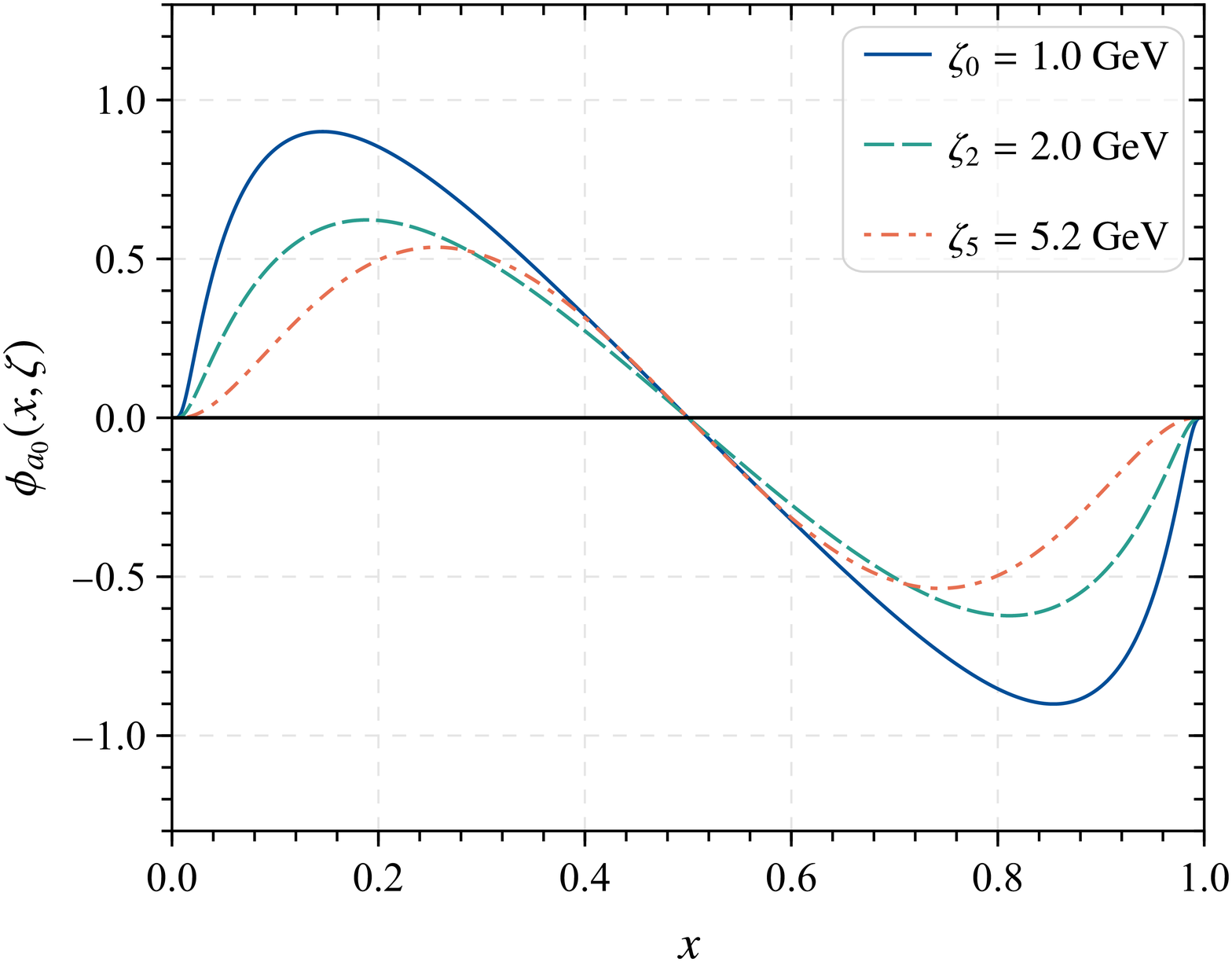}
\caption{The $a_0(980)$-meson leading-twist LCDA $\phi_{a_0}(x,\zeta)$ changed with three different scales $\zeta = (1.0,2.0,5.2)~{\rm GeV}$.}
\label{Fig:DA}
\end{figure}

\begin{table}[t]
\centering
\renewcommand\arraystretch{1.3}
\small
\caption{The LCHO model parameters $A_{a_0}$ (in unit: GeV$^{-1}$), $\beta_{a_0}$ (in unit: GeV), $\alpha_{a_0}$ and goodness of fit $P_{\chi^2_{\rm min}}$ changed with the factorization scale $\zeta=(1.0,2.0,5.2)~{\rm GeV}$.}
\begin{tabular}{ l  l  l  l l}
\hline
$\zeta$~~~~~~~~~&$~A_{a_0}$~~~~~~~~~~~&$\beta_{a_0}$~~~~~~&$\alpha_{a_0}~~~~~~~$ & $P_{\chi^2_{\rm min}}$\\
\hline
$1.0$ &  $-260$  & $0.5$  &  $-0.43$  &  $0.773$\\
$2.0$ &  $-380$  & $0.5$  &  $-0.05$   &  $0.833$\\
$5.2$ &  $-1580$ & $0.5$  &  $~~0.85$   &  $0.129$\\
\hline
\end{tabular}
\label{parameters}
\end{table}

Firstly, we list the $a_0(980)$-meson leading-twist LCDA $\xi$-moments $\langle\xi^n_{a_0}\rangle_\zeta$ with three different scales $\zeta = (1.0,2.0,5.2)~{\rm GeV}$ in Table~\ref{Tab:moments}. In which, the accurate of our calculation is up to 9th-order. It can be seen that the absolute value of $\xi$-moments decreases as the scale $\zeta$ increases. Secondly, the absolute value of $\xi$-moments decreases with the $n$ increases, which shows that our calculation has good convergence. Then, we adopt the least squares method to fit the $\xi$-moments $\langle\xi^n_{a_0}\rangle_\zeta$. Next, the fitting model parameters with different scale $\zeta$ are given in Table~\ref{parameters}. Based on the experience of other mesons, we take the WF model parameter $\beta_{a_0} = 0.5$. Obviously, $A_{a_0}$ gradually decreases with the increment of the scale $\zeta$. However, the goodness of fit $P_{\chi^2_{\rm min}}$ increases firstly and then decreases.

\begin{figure}[t]
\center
\includegraphics[width=0.45\textwidth]{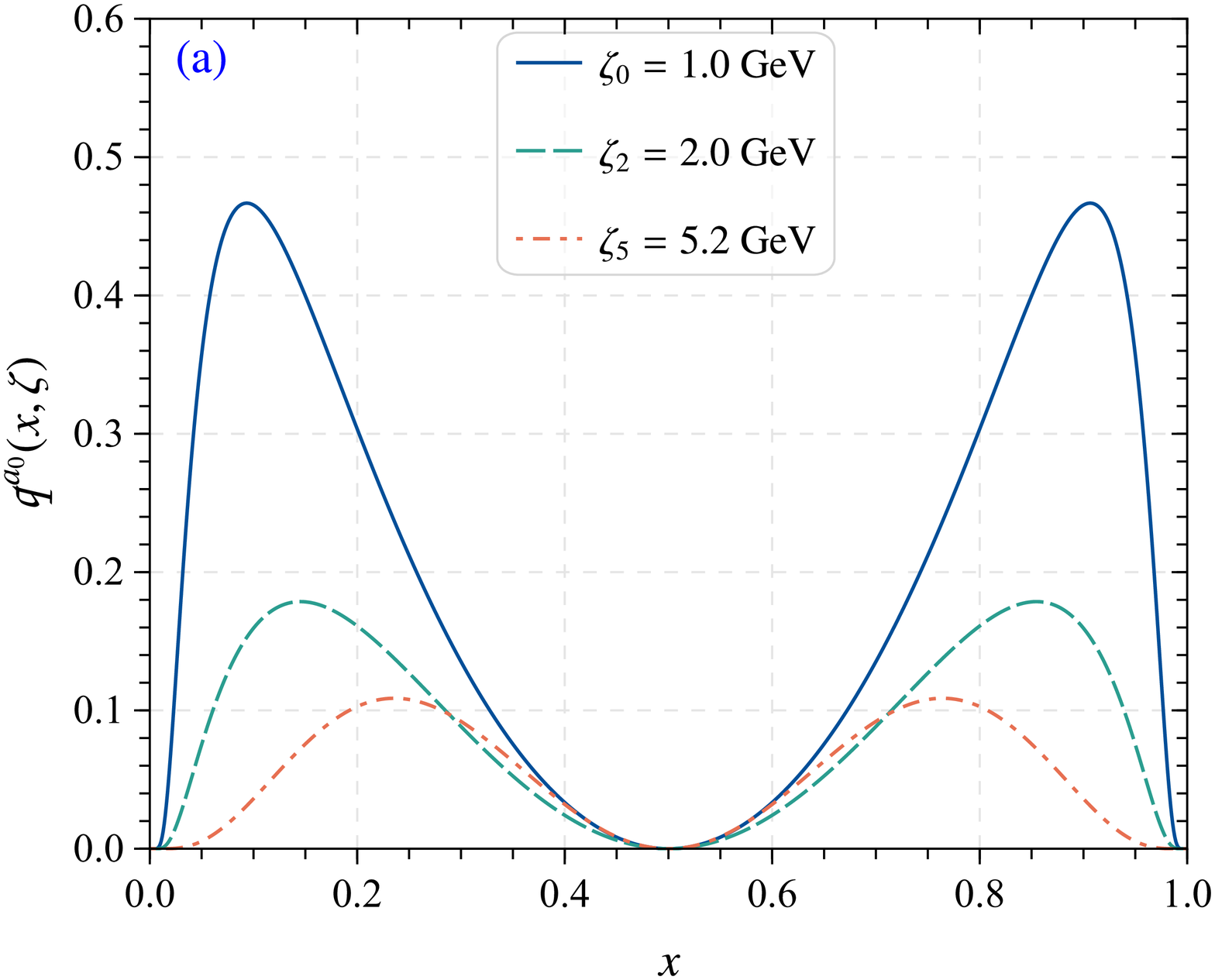}
\includegraphics[width=0.45\textwidth]{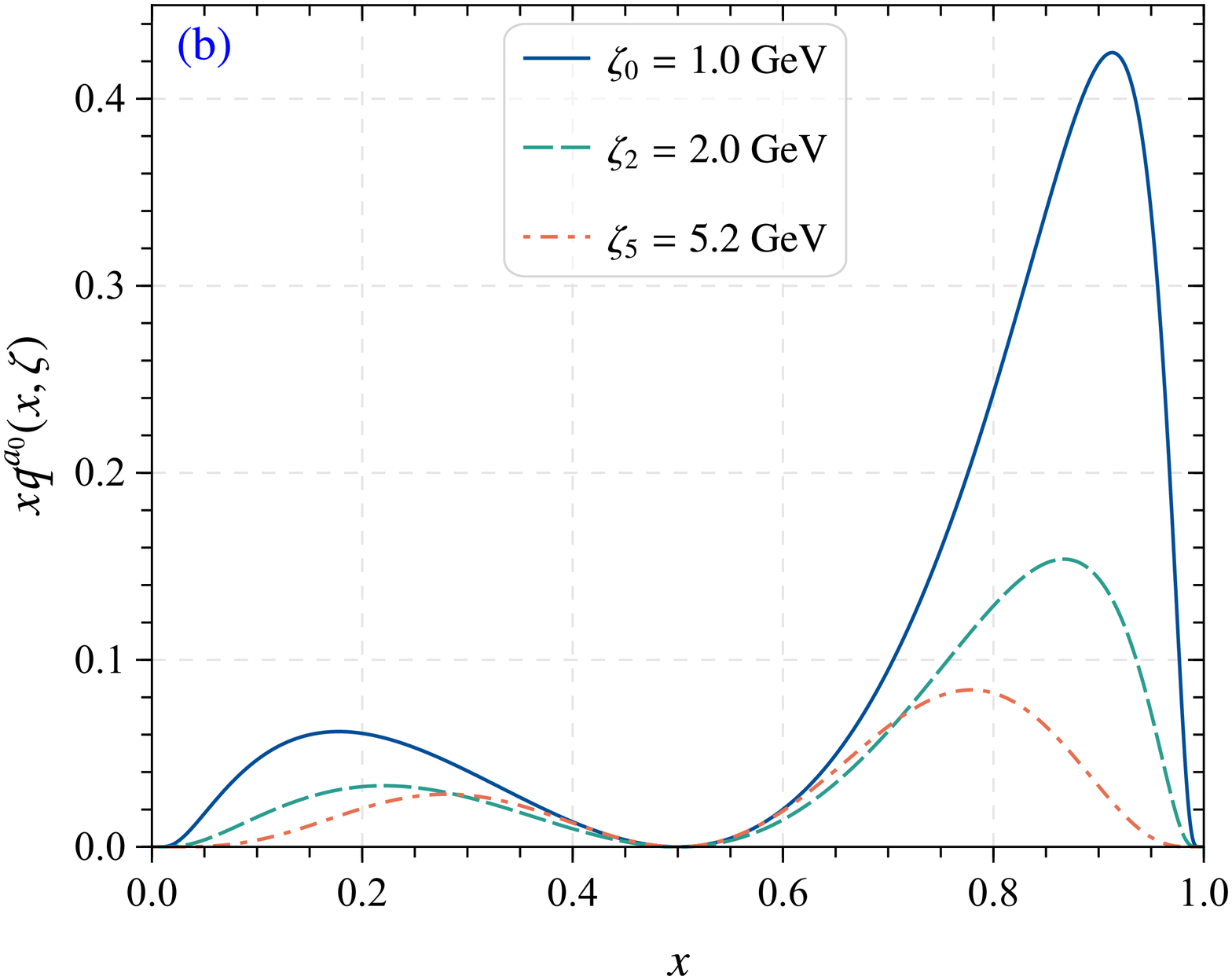}
\caption{(Color online) The $a_0(980)$-meson valence quark distribution function $\q(x,\zeta)$ and $x\q(x,\zeta)$ with different scales.}
\label{fig:PDFs}
\end{figure}

With the resultant LCHO model parameters, the curves of $a_0(980)$-meson leading-twist LCDA with three scales $\zeta$ are shown in Fig.~\ref{Fig:DA}. The figure shows that
\begin{itemize}
    \item The behaivor of the three curves are tend to antisymmetric, which will equal to zero when the LCDA integrate with respect to $x$, e.g.
        \begin{eqnarray}
            \int_0^1 dx \phi_{a_0} (x,\zeta) = 0.
        \end{eqnarray}
        Meanwhile, the three curves go through the zero at the location $x = 0.5$.
    \item The absolute value of the peaks are decreased with the increase of $\zeta$ and the $x$-location of the peaks are tend to 0.5 with the $\zeta$ increase. When the scale tend to infinity i.e. $\zeta \to \infty$, the curve of $a_0(980)$-meson LCDA will tend to asymptotic form $\phi_{a_0} (x,\infty) = 0$
\end{itemize}

Secondly, after taking the LCHO parameters into the $a_0(980)$-meson valence-quark distribution function, e.g.  Eq.~\eqref{PDF_model}, the predictions of $\q(x,\zeta)$ can be obtained. The curves of $a_0(980)$-meson valence-quark distribution function $\q(x,\zeta)$ and $x\q(x,\zeta)$ with different scales $\zeta$ are shown in Fig.~\ref{fig:PDFs}, which shows that the value of peaks decrease with the increase of scale $\zeta$. Since the $a_0(980)$-meson leading-twist LCDA is antisymmetric behavior under $u\to (1-u)$ interchange in the SU$_f$(3) limit, its valence-quark distribution function $x\q(x,\zeta)$ is tend to zero at $x=0.5$. Additionally, valence-quark distribution function tends to bimodal behavior. In general, the valence-quark distribution functions of pion and kaon tend to a unimodal behavior \cite{Cui:2020tdf}.

\begin{table}[t]
\centering
\renewcommand\arraystretch{1.3}
\small
\caption{The Mellin moments for $a_0(980)$-meson leading-twist distribution function $\langle x^n \mathpzc q^{a_0}\rangle_\zeta$ with different scales $\zeta=(1.0, 2.0, 5.2)~\rm{GeV}$.}
\begin{tabular}{ l  c  c  c c}
\hline
$\zeta$~~~~~~& ~~~~~~$\langle x \mathpzc q^{a_0}\rangle_\zeta$~~~~~~&~~~~~~$\langle x^2 \mathpzc q^{a_0}\rangle_\zeta$~~~~~~&~~~~~~$\langle x^3 \mathpzc q^{a_0}\rangle_\zeta$ ~~~~\\
\hline
$1.0$~~ & ~~$0.100$~~ & ~~$0.074$~~& ~~$0.062$~~~~\\
$2.0$~~ & ~~$0.044$~~ & ~~$0.031$~~& ~~$0.024$~~~~\\
$5.2$~~ & ~~$0.026$~~ &~~$0.017$~~& ~~$0.012$~~~~\\
\hline
\end{tabular}
\label{Tab:moments}
\end{table}
\begin{figure}[t]
\centering
\includegraphics[width=0.45\textwidth]{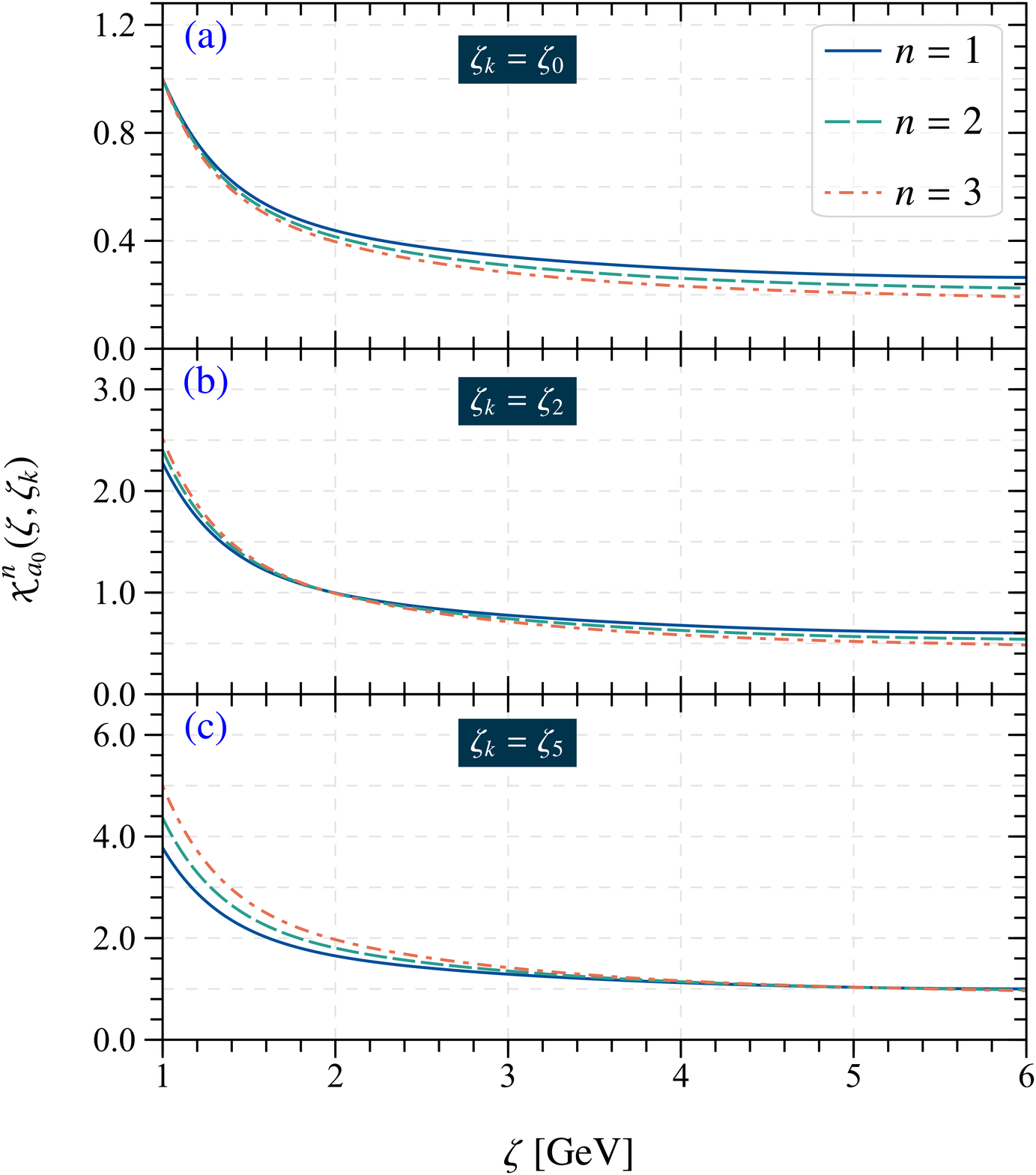}
\caption{(Color online) The predicted ratio of Mellin moments of the $a_0(980)$-meson valence-quark distribution function $\mathpzc x_{\,a_0}^n(\zeta,\zeta_k)$ with three fixed scales $\zeta_k = (1.0,2.0,5.2)~{\rm GeV}$ changed with arbitrary scales $\zeta$ in the range $\zeta=[1,6]~{\rm GeV}$. In which, the $n$ is taken as $n = (1,2,3)$ respectively.}
\label{fig:xin ratio}
\end{figure}

Using the meson's valence quark distribution function, we can get the Mellin moments $\langle x^n \mathpzc q^{a_0}\rangle_{\zeta}$ of the $a_0(980)$-meson valence-quark distribution function, which are presented in Table~\ref{Tab:moments}. From the table, we can see that the Mellin moments are convergence with the order $n$ increased. Meanwhile, the Mellin moments are also convergence with the scale $\zeta$ increased. This agrees with the Mellin moments of pion's valence-quark distribution function decreases with the increase of scale. The greater $\zeta$, the smaller the value of the moments in Refs.~\cite{Cui:2020tdf}. It proves our prediction of the Mellin moments $\langle x^n \mathpzc q^{a_0}\rangle_\zeta$ are reasonable.

Finally, we also calculate the ratio $\mathpzc x^n_{\,a_0}(\zeta,\zeta_k)$ of Mellin moments changed with the scale $\zeta$. The predictions of the ratio of Mellin moments with three fixed scales $\zeta_k = (1.0, 2.0, 5.2)~{\rm GeV}$ and different order $n = (1,2,3)$ are depicted in Fig.~\ref{fig:xin ratio}. From which the $\mathpzc x^n_{\,a_0}(\zeta,\zeta_k)$ are increased with index $n$ before the point of $\zeta_k$, and decreased with $n$ after $\zeta_k$. The curves of $\mathpzc x^n_{\,a_0}(\zeta,\zeta_k)$ are decreasing with the $\zeta$ increases. The curves will coincide with each other when the scale $\zeta$ and $\zeta_k$ tend to infinity.

\textit{Summary} --- In this paper, we fit moments $\langle x^n_{a_0}\rangle_\zeta$ with the least squares method to obtain the free model parameters $A_{a_0}$, $\beta_{a_0}$ and $\alpha_{a_0}$ at the scales $\zeta = (1.0,2.0,5.2)~{\rm GeV}$. Meanwhile, the goodness of fit $P_{\chi^2_{\rm min}}$ is also given. Then, we present the curves of $a_0(980)$-meson leading-twist LCDA shown in Fig.~\ref{Fig:DA}. After constructing the relationship between $a_0(980)$-meson leading-twist WF and PDFs, the $a_0(980)$-meson valence quark distribution function $\q(x,\zeta)$ and $x\q(x,\zeta)$ with different scales are shown in Fig.~\ref{fig:PDFs}, which tends to bimodal behavior. The LCDA and PDFs are tend to zero at the location $x = 0.5$ due to the antisymmetry of the WF. Based on the $a_0(980)$-meson valence quark distribution function, we can get the first three order Mellin moments $\langle x \q\rangle_\zeta$ of the $a_0(980)$-meson valence quark DF shown in Table~\ref{Tab:moments}. Referring to the predicted pion's Mellin moments, our predicted result is quite reasonable. At the same time, we also give the ratio $\mathpzc x^n_{\,a_0}(\zeta,\zeta_k)$ of Mellin moments with $\zeta= (1.0, 2.0, 5.2)~ {\rm GeV}$ shown in Fig.~\ref{fig:xin ratio}. The ratio $\mathpzc x^n_{\,a_0}(\zeta,\zeta_k)$ shows a downward tendency with the increase of $\zeta$.
\\

\textit{Acknowledgments} --- This work was supported in part by the National Natural Science Foundation of China under Grant No.12265010, No.12265009, the Project of Guizhou Provincial Department of Science and Technology under Grant No.ZK[2021]024, the Project of Guizhou Provincial Department of Education under Grant No.KY[2021]030.

\end{document}